\documentclass[10pt,a4paper,twocolumn,notitlepage,prd,showpacs,preprintnumbers,superscriptaddress,nofootinbib]{revtex4-1}
\usepackage{float}
\usepackage{amsmath,amssymb}
\usepackage{cancel}
\usepackage{graphicx}

\makeatletter

\usepackage{bm}
\usepackage{tensor}
\usepackage{comment}
\usepackage{ifpdf}
\usepackage{slashed}
\usepackage{color}
\usepackage[mathscr]{eucal}

\ifpdf
  \usepackage{graphicx}     
  \usepackage[bookmarksopen,colorlinks=true,linkcolor=bblue,citecolor=bblue,urlcolor=ppink]{hyperref}
\else     
  \usepackage[dvipdfmx]{}     
  \usepackage[dvipdfmx,bookmarksopen,colorlinks=true,linkcolor=bblue,citecolor=ppink,urlcolor=darkred]{hyperref}
\fi

\definecolor{red}{rgb}{1,0,0}
\definecolor{darkred}{rgb}{0.6,0,0}
\definecolor{darkgreen}{rgb}{0.992447,0.623778,0.034597}
\definecolor{ppink}{rgb}{1,0.4,0.4}
\definecolor{bblue}{rgb}{0.284602,0.317763,0.963947}

\newcommand{\Mpl}{M_{\rm Pl}}

\def\Mpl{M_{\rm Pl}}

\newcommand{\ltsim}{\protect\raisebox{-0.5ex}{$\:\stackrel{\textstyle <}{\sim}\:$}}

\newcommand{\footnoteref}[1]{\protected@xdef\@thefnmark{\ref{#1}}\@footnotemark}

\makeatother

\begin{document}

\preprint{IPMU17-0144}

\title{Inflaton fragmentation in E-models of cosmological $\alpha$-attractors}

\author{Fuminori Hasegawa}
\email[fuminori@icrr.u-tokyo.ac.jp]{}
\affiliation{ICRR, University of Tokyo, Kashiwa, 277-8582, Japan}

\affiliation{Kavli IPMU (WPI), UTIAS, University of Tokyo, Kashiwa, 277-8583,
Japan}

\author{Jeong-Pyong Hong}
\email[hjp0731@icrr.u-tokyo.ac.jp]{}
\affiliation{ICRR, University of Tokyo, Kashiwa, 277-8582, Japan}

\affiliation{Kavli IPMU (WPI), UTIAS, University of Tokyo, Kashiwa, 277-8583,
Japan}
\begin{abstract}
\noindent  Cosmological $\alpha$-attractors are observationally favored due to the asymptotic flatness of the potential. 
Since its flatness induces the negative pressure, the coherent oscillation of the inflaton field could fragment into quasi-stable localized objects called I-balls (or ``oscillons"). 
We investigated the possibility of I-ball formation in E-models of $\alpha$-attractors. 
Using the linear analysis and the lattice simulations, we found that the instability sufficiently grows against the cosmic expansion and the inflaton actually fragments into the I-balls for $\alpha\ltsim10^{-3}$.  \end{abstract}
\maketitle
\section{Introduction
\label{sec:introduction}}
The inflationary cosmology~\cite{Starobinsky:1980te,Guth:1980zm,Linde:1981mu,Linde:1983gd} solves many puzzles of the standard big bang cosmology, such as the flatness and horizon problems. While various inflation models have been proposed, recent observations of cosmic microwave background~(CMB)~\cite{Ade:2015lrj} have excluded many classes of models, including single power-law potential models, through the combined constraints on the spectral index $n_s$ and the tensor-to-scalar ratio $r$.

A new class of inflationary models that is consistent with the observations, was recently proposed~\cite{Kallosh:2013hoa,Galante:2014ifa,Kallosh:2015lwa}. 
These models, called $\alpha$-attractors, are constructed in a quite bottom-up manner, which include various types of inflation models unified by only one (phenomenologically) free-parameter $\alpha$. 
For example, the quadratic inflation model~\cite{Linde:1981mu,Linde:1983gd} and the Starobinsky model~\cite{Starobinsky:1980te,Mukhanov:1981xt} are reproduced by $\alpha=\infty$ and $\alpha=1$, respectively.

The $\alpha$-attractors are categorized into two subclasses: T-models and E-models, which are characterized by the following potentials, 
\begin{align}
V_{\rm T}(\phi)=V_0\tanh^{2n}\left(\frac\phi{\sqrt{6\alpha}M_{\rm pl}}\right),\\
V_{\rm E}(\phi)=V_0\left(1-e^{-\sqrt{\frac2{3\alpha}}\frac{\phi}{M_{\rm pl}}}
\right)^{2n},
\end{align}
respectively.

It is known that if the real scalar field $\phi$ oscillates with the potential shallower than quadratic, $\phi$ fragments into the quasi-stable lumps, which are called I-balls (or oscillons)~\cite{Copeland:1995fq,Kasuya:2002zs,Amin:2010dc,Gleiser:2011xj}. 
Since the oscillation in the potential shallower than quadratic induces the negative pressure (e.g. the scalar potential $V(\phi)=\phi^{2-K}$ leads to the equation of state $p=-\frac{K}{4}\rho$, where $0<K\ll1$), the real scalar field tends to form a localized bound state which is energetically favored. 
The formation of the I-ball has been reported in various type of the potential such as a double-wall potential~\cite{Gleiser:1993pt,Fodor:2006zs}, the axion-like potential~\cite{Kolb:1993hw}, or the potential of the string moduli~\cite{Antusch:2017flz}.
The I-ball is defined as the scalar configuration which minimizes the energy with a fixed adiabatic charge $I$ and its conservation guarantees the longevity of the I-ball~\cite{Kasuya:2002zs}. 
It is known, however, since the adiabatic charge is conserved only approximately, the I-ball is quasi-stable and will decay in a finite time. Recently, analytic estimation of the lifetime of the I-balls are performed in classical level~\cite{Mukaida:2016hwd}, which is in good agreement with the numerical simulation. 

The fragmentation of the inflaton field into such quasi-stable lumps can have an impact on the cosmology after inflation. 
There have been a lot of studies about the case the inflaton forms the I-ball rather than oscillates coherently \cite{McDonald:2001iv}. 
For instance, the I-ball formation alters the reheating process because the inflaton energy is transferred to the light particles through the decay of the I-balls. 
Therefore, the reheating temperature would be different from the perturbative one.
The number of $e$-foldings at CMB pivot scale, $N_*$, is also changed, since it is related to the reheating temperature. 
The production of the gravitational waves due to the I-ball formation is also discussed~\cite{Antusch:2016con,Liu:2017hua}.

In this paper, we focus on the I-ball formation in the E-models of the $\alpha$-attractors.
There have been some studies on the I-ball formation in the $\alpha$-attractors and it was pointed out that the I-balls are formed in the case of T-models~\cite{Lozanov:2016hid,Kim:2017duj}. 
They find that in order for the inflaton to fragment into the I-balls, $~\alpha\lesssim\alpha_{\rm th}^{\rm T}\sim10^{-4}$ is required for $n=1$. 
In the T-model, the negative quartic coupling of the inflaton plays a crucial role to flatten the inflaton potential. 
In the case of E-models, however, the sign of the quartic coupling is positive and it does not flatten the inflaton potential. 
Instead of the quartic coupling, the negative cubic term, which is absent in the T-model, flattens the potential. 
Since the negative cubic term is asymmetric and flatten the potential only for the region $\phi>0$, the condition or the property of the I-ball formation can be different from the case of the T-models. 
To see whether I-ball is formed, we first simulate the growth of the linear fluctuations of the inflaton $\phi$ for given $\alpha$. 
We find that the fluctuations become larger than the background value against the cosmic expansion only for $\alpha\lesssim\alpha_{\rm th}^{\rm E}\sim10^{-3}$. 
We also perform the lattice simulations in 1D, 2D and 3D to follow the dynamics even in non-linear regime. 
Consequently, we confirm the I-ball is actually formed for $\alpha\lesssim10^{-3}$.
This result is consistent with the absence of the I-ball formation in the Starobinsky model, which corresponds to the E-model with $(n,\alpha)=(1,1)$~\cite{Takeda:2014qma}

The rest of the paper is organized as follows. 
In Sec.~\ref{sec:emi}, we review the E-models of the $\alpha$-attractors and discuss their observables. 
In Sec.~\ref{sec:bp}, we analytically derive the I-ball solution in the E-models. 
In Sec.~\ref{sec:dr}, we examine the growth of the instability band for small $\alpha$ in E-models. 
In Sec.~\ref{sec:nu}, we present the results of lattice simulations of I-ball formation in E-models. 
In Sec.~\ref{sec:cosmo}, we discuss the effect of I-ball formation on the cosmology. Sec.~\ref{sec:conc} is devoted to the conclusions.
\section{E-models of the $\alpha$-attractors}
\label{sec:emi}
The recently proposed $\alpha$-attractors are consistent with the observations due to the flatness of the potential in the large field regime $\phi/\sqrt{\alpha}\gg1$. One can regard the origin of its flatness as the pole in the kinetic term e.g,
\begin{align}
\mathcal{L}\supset \frac{3\alpha(\partial\tau)^2}{4\tau^2}-V(\tau),
\end{align}
where $\tau$ denotes a real scalar field.
Once we rewrite the potential by the canonically normalized field $\phi$, which is related to $\tau$ as 
\begin{align}
\tau={\rm exp}\left(-{\sqrt{\frac{2}{3\alpha}}\frac{\phi}{M_{\rm pl}}}\right), 
\end{align}
the large field region of the potential is exponentially stretched and becomes asymptotically flat.
The $\alpha$-attractors are naturally embedded into the $\mathcal{N}=1$ supergravity exploiting the hyperbolic geometry of the Poincar\'{e} disk or half-plane~\cite{Carrasco:2015uma}. Then only one parameter $\alpha$ is related to the curvature of the K\"{a}hler geometry $R_{K}=-\frac{2}{3\alpha}$, which is also phenomenologically arbitrary parameter.

\begin{figure}[t]
\centering
  \includegraphics[width=1\linewidth]{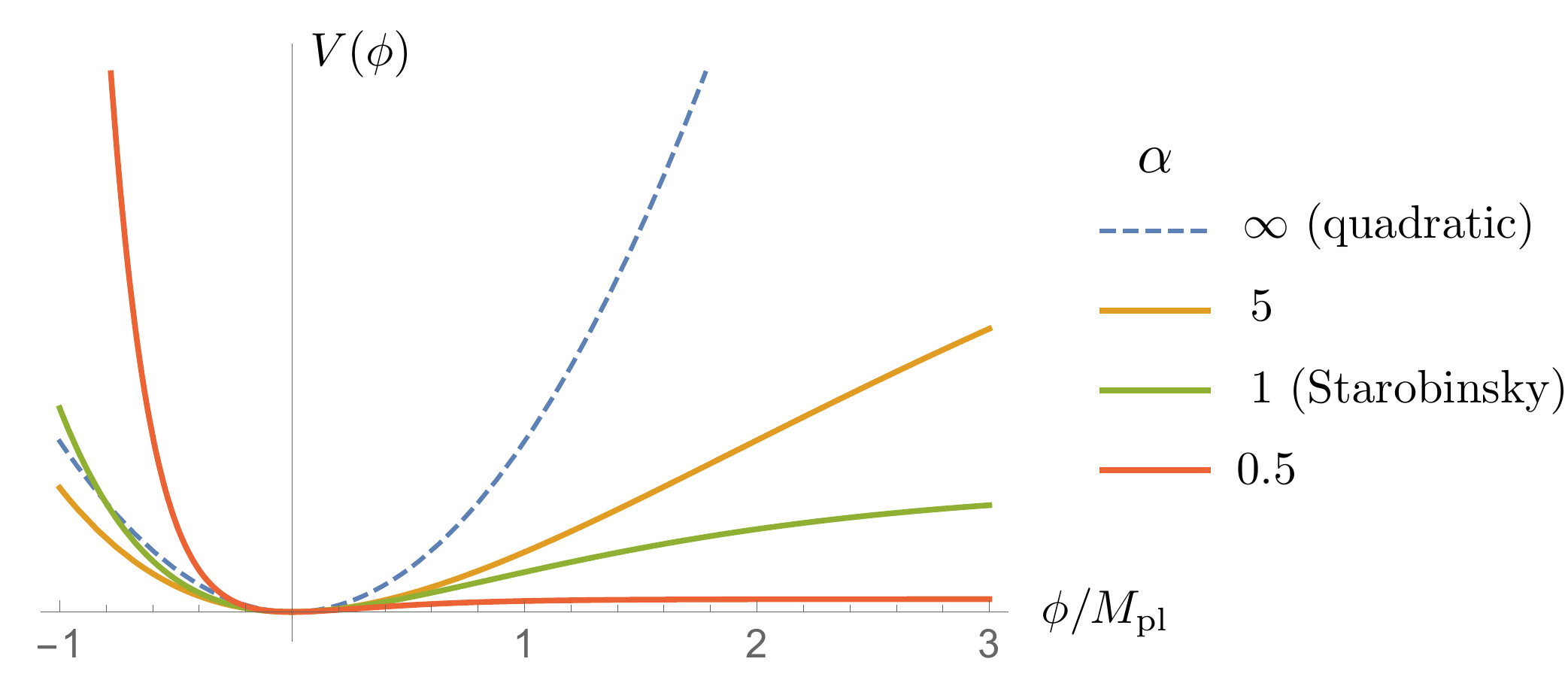}
 \caption{The behavior of the E-model potentials~$(n=1)$ for different choices of $\alpha$. The case of $\alpha=1$ corresponds to the Starobinsky model.}
\label{fig:emodel}
\end{figure}

The $\alpha$-attractors can be categorized into two subclasses: T-models and E-models.
The E-models are specified by the following asymmetric potential~\cite{Carrasco:2015pla,Carrasco:2015rva},
\begin{align}
V_{\rm E}(\phi)=\frac34m^2M_{\rm pl}^2\alpha\left(1-e^{-\sqrt{\frac2{3\alpha}}\frac{\phi}{M_{\rm pl}}}\right)^{2n},
\end{align}
which becomes asymptotically flat for $\phi\gtrsim\sqrt{\alpha}M_{\rm pl}$.
One can see that the choice $(n,\alpha)=(1,1)$ reproduces the shape of the Starobinsky potential which originally described by scalaron mode in the $R+R^2$ gravity~\cite{Starobinsky:1980te}.  
In Fig.~\ref{fig:emodel}, we present the behavior of the potential for different choices of $\alpha$ in the case of $n=1$. Since the lump can be quasi-stable only in the case the potential is quadratic around the origin, we consider $n=1$ in the following section\footnote{The localized solutions also exist for $n\neq1$ called ``transients". They immediately decay via self-interaction after their formation~\cite{Lozanov:2016hid}. }.
\subsection{Inflationary observables}
\begin{figure}[t]
\centering
  \includegraphics[width=1\linewidth]{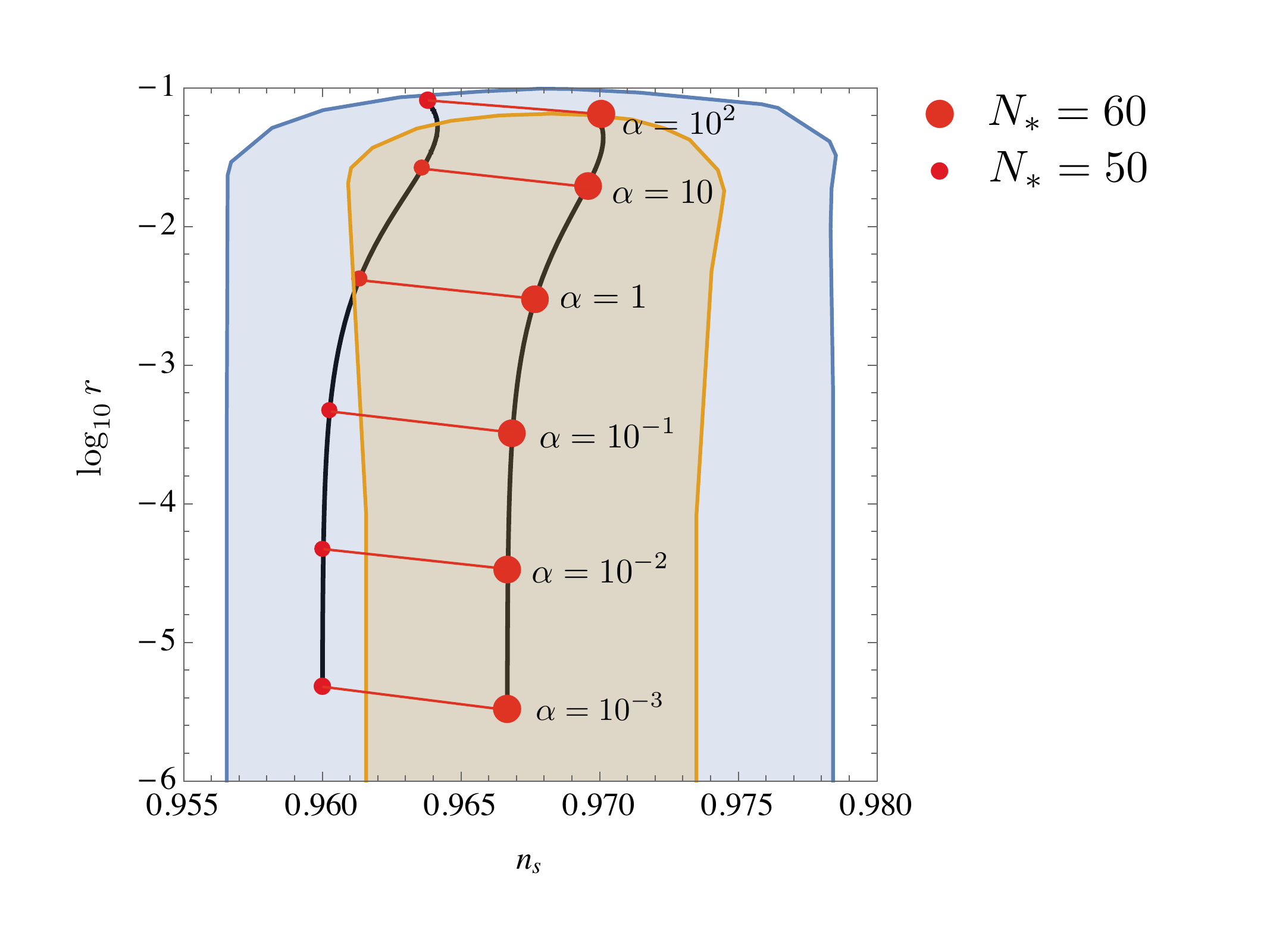}
 \caption{The prediction of the E-model~$(n=1)$ for various $\alpha$. The case of $\alpha=1$ corresponds to the Starobinsky model. }
\label{fig:nsr}
\end{figure}

In this section, we discuss the inflationary observables predicted by the $\alpha$-attractors. 
The amplitude of primordial curvature perturbation are given by
\begin{align}
\label{amp}
\mathcal{P}_\zeta&\equiv\frac{V}{24\pi^2\epsilon M_{\rm pl}^4}\simeq\frac{N_*^2m^2}{24\pi^2 M_{\rm pl}^2},
\end{align}
and the spectral index of primordial curvature perturbation and the tensor to scalar ratio are given by
\begin{align}
\label{ns}n_s&\equiv1-6\epsilon+2\eta\simeq1-\frac2N_*,\\
\label{r}r&\equiv16\epsilon\simeq\frac{12\alpha}{N_*^2}.
\end{align}
All the quantities are evaluated at $N_*$ when the CMB scale exits the horizon and valid for $\alpha\ll1$.
According to the latest Planck results~\cite{Ade:2015lrj}, the amplitude of the primordial curvature perturbation at the CMB scale $k_\ast\simeq0.002~{\rm Mpc}^{-1}$ is given as
\begin{align}
\mathcal{P}_\zeta(k_\ast)\simeq2.4\times10^{-9}.
\end{align}
Then, the inflaton mass $m$ is determined as
\begin{align}
m\simeq1.4\times10^{-5}M_{\rm pl}\left(\frac{N_\ast}{55}\right)^{-1}.
\label{eq:mas}
\end{align}
The spectral index and the tensor-to-scalar ratio are also constrained by the latest Planck results as
\begin{align}
n_s=0.9677\pm0.006,~r<0.07~{\rm (Planck~2015)},
\end{align}
which are in good agreement with the predictions of the E-models: 
\begin{align}
n_s&\simeq0.964,~r=4\alpha\times10^{-3},~(N_*=55)
\end{align}
for $\alpha\ltsim\mathcal{O}(10)$. We plot the prediction of the E-models for various $\alpha$ in $(n_s,~r)$ plane, in Fig.~\ref{fig:nsr}.

Now we also comment on the inflationary observables predicted by T-models of the $\alpha$-attractors. 
The potential shape of the T-model is very similar to that of the E-models in large field regime. 
In fact, their predictions of the inflationary observables coincide with each other for $\alpha\ll N$, that is, two models are almost degenerate in $(n_s,~r)$ plane. 
However, the condition for the formation of I-balls in E-models differs from that in T-models, as we will see in the following sections.

Finally, we calculate the field value of the inflaton at the end of the inflation $\phi_{\rm end}$, which we will use later. The end of the inflation is defined by the condition $\epsilon,|\eta|\sim1$.
At that time, the field value of the inflaton takes a value
\begin{align}\label{pe}
\phi_{\rm end}\simeq\sqrt{\frac{3\alpha}2}M_{\rm pl}\log\left[\sqrt{\frac4{3\alpha}}+1\right].
\end{align}
In the following, we adopt $\phi_{\rm end}$ as the initial value of the oscillation after inflation. 
Strictly speaking, of course, the amplitude of the oscillation is slightly larger than $\phi_{\rm end}$ since $\dot{\phi}_{\rm end}\neq0$. 
However, it damps to $\phi_{\rm end}$ by cosmic expansion in a time scale that is negligible compared to that of the fragmentation of the inflaton. 
Since the fragmentation is more efficient for a larger oscillation amplitude, this choice of the initial condition is rather conservative for the I-ball formation.
\section{The I-ball solution and its profile}
\label{sec:bp}
When the dynamics is nearly periodic, the area of the track in the phase space is conserved, which is called adiabatic invariant $I$:
\begin{align}
I\equiv\frac1{2\omega}\int d^3x\overline{\dot{\phi^2}},
\end{align}
where $\omega$ is the frequency of the periodic motion, or oscillation.
I-ball profiles are defined as the solutions which minimize the one-period averaged energy, with fixed $I$. 
To find them, we can use the Lagrange multiplier method. 
However, in the E-models, while the dynamics after inflation is dominated by harmonic oscillation, cubic correction in the potential make the oscillation asymmetric and averaging the energy rather involved. 
Thus, here we derive the profile of the I-ball using an alternative method called $\epsilon$-expansion~\cite{Farhi:2007wj,Fodor:2009kf,Dashen:1975hd}, where we directly solve the time and space dependence of the oscillating solution in the small-amplitude approximation. 
Although the physical relation to the $I$ conservation is not clear, this method describes the I-ball profile very well\footnote{We note the $\epsilon$-expansion is not applicable for non-polynomial potentials like $V\sim \phi^{2-K},~0<K\ll1$, which has an I-ball solution}. 
Since the analytic solutions exist only for 1+1 dimension, we will perform the analysis in 1+1 dimension.
Then, the equation of motion of $\phi$ is given as
\begin{align}
\frac{d^2\phi}{dt^2}-\frac{d^2\phi}{dx^2}+V'(\phi)=0,
\label{eq:eqep}
\end{align}
where
\begin{align}
V'(\phi)&=\sqrt{\frac{3\alpha}2}m^2M_{\rm pl}e^{-\sqrt{\frac2{3\alpha}}\frac{\phi}{M_{\rm pl}}}\left(1-e^{-\sqrt{\frac2{3\alpha}}\frac{\phi}{M_{\rm pl}}}\right).
\label{eq:forc}
\end{align}
Since the I-ball solution is dominated by the harmonic oscillation, the amplitude of the harmonic oscillation $\Phi(x)$ at the center of the I-ball $(x=0)$ must satisfy 
\begin{align}
\epsilon\equiv\frac{\Phi(0)}{\sqrt{\alpha}M_{\rm pl}}\lesssim1,
\end{align}
here we define the small parameter $\epsilon$ and use this quantity as a parameter of the expansion. 
The inflaton field $\phi(x,~t)$ is then expanded as
\begin{align}
 \frac{\phi}{\sqrt{\alpha}M_{\rm pl}}\equiv\epsilon\phi_1+\epsilon^2\phi_2+\epsilon^3\phi_3+\mathcal{O}(\epsilon^4).
 \end{align}
Although we can obtain the solution by substituting it into Eq.~(\ref{eq:eqep}), we introduce following variables in order to simplify the calculation:
\begin{align}
\tau&\equiv mt\sqrt{1-A\epsilon-B\epsilon^2+\mathcal{O}(\epsilon^3)},\\
\chi&\equiv mx\left(\epsilon+\mathcal{O}(\epsilon^2)\right),
\end{align}
here we omit the $\mathcal{O}(\epsilon^0)$ term in $\chi$ according to the fact that the solution losses the $\epsilon$-dependence for the harmonic limit $\epsilon\rightarrow0$.
Then, the equation of motion Eq.~(\ref{eq:eqep}) is decomposed as
 \begin{align}
\phi_{1\tau\tau}+\phi_1&=0,\label{eq:phi1}\\
\phi_{2\tau\tau}+\phi_2&=A\phi_{1\tau\tau}+\frac3{\sqrt{6}}\phi_1^2,\label{eq:phi2}\\
\phi_{3\tau\tau}+\phi_3&=B\phi_{1\tau\tau}+\phi_{1\chi\chi}+A\phi_{2\tau\tau}+\sqrt{6}\phi_1\phi_2-\frac79\phi_1^3,\label{eq:phi3}
\end{align}
where we expanded the Eq.~(\ref{eq:forc}) as well. The solution for the Eq.~(\ref{eq:phi1}) is easily obtained as
\begin{align}
\phi_1(\chi,\tau)\equiv f(\chi)\cos(\tau),
\end{align}
which reproduces the coherent oscillation for $\epsilon\rightarrow0$ limit.
Then, the Eqs.~(\ref{eq:phi2}), (\ref{eq:phi3}) reduce to
\begin{align}
\phi_{2\tau\tau}+\phi_2=&-Af\cos(\tau)+\frac3{\sqrt{6}}f^2\cos^2(\tau),\label{eq:phi2s}\\
\phi_{3\tau\tau}+\phi_3=&-Bf\cos(\tau)+f_{\chi\chi}\cos(\tau)\\
&+A\phi_{2\tau\tau}+\sqrt{6}f\cos(\tau)\phi_2-\frac79f^3\cos^3(\tau).\label{eq:phi3s}
\end{align}
In order to obtain the stable solution, we eliminate the secular term which diverges for $\tau\rightarrow\infty$ in Eq.~(\ref{eq:phi2s}). By choosing $A=0$ we can obtain the stable solution for $\phi_2$ as
\begin{align}
\phi_2=f^2\left[\frac{\sqrt{6}}4-\frac{\sqrt{6}}{12}\cos(2\tau)\right].
\end{align}
Then, the last equation (\ref{eq:phi3s}) reduce to
\begin{align}
\phi_{3\tau\tau}+\phi_3=\left[f_{\chi\chi}-Bf+\frac23f^3\right]\cos(\tau)-\frac49f^3\cos(3\tau).
\label{eq:phi3ss}
\end{align}
Again, by the stability of the solution, the first term must vanish:
\begin{align}
f_{\chi\chi}-Bf+\frac23f^3=0.
\label{eq:BB}
\end{align}
This determines the spatial profile of the solution, which is analytically given as follows.
\begin{align}
f(\chi)=f(0){\rm sech}\left[\frac1{\sqrt{3}}f(0)\chi\right].
\label{eq:ana}
\end{align}
Since $\phi_1$ represents nothing but the harmonic mode of the oscillation, 
\begin{align}
\sqrt{\alpha}\Mpl\epsilon f(\chi)=\Phi(\chi)
\end{align}
should be satisfied by definition. 
Then, the amplitude of the harmonic oscillation $\Phi(\chi)$ is the represented as
\begin{align}
\Phi(x)&\equiv\Phi(0){\rm sech}\left[\frac{\Phi(0)}{\sqrt{3\alpha}M_{\rm pl}}mx\right]
\end{align}
Finally, solving Eq.~(\ref{eq:phi3ss}), we obtain the following perturbative solution up to the order of $\epsilon^3$ as
\begin{align}
\phi(x,t)\simeq\Phi(x)&\left[\cos(\tau)+\frac{\sqrt{6}}{12}\left(\frac{\Phi(x)}{\sqrt{\alpha}M_{\rm pl}}\right)(3-\cos(2\tau))\right.\nonumber\\
&\left.+\frac1{18}\left(\frac{\Phi(x)}{\sqrt{\alpha}M_{\rm pl}}\right)^2\cos(3\tau)\right],
\label{eq:ana2}
\end{align}
where $\tau$ is given as 
\begin{align}
\tau=\sqrt{1-\frac1{3}\left(\frac{\Phi(0)}{\sqrt{\alpha}M_{\rm pl}}\right)^2}mt .
\end{align}
We note that the asymmetric property appears as a second order correction, which induces the instability as we will see in the next section.
We compare these analytic profiles with those obtained by 1D lattice simulations in Sec.~\ref{sec:nu}. 
\section{Growth of the instability}
\label{sec:dr}
As we have seen in the previous section, the potential of E-model is asymmetric due to the cubic term, which makes it flatter than quadratic only for $\phi>0$. 
This flatness allows the existence of the quasi-stable lump solution of the inflaton during the oscillation era.
However, the existence of the solution itself does not guarantee the I-ball formation in the expanding universe.
For example, in the Starobinsky model, inflaton does not fragment into the lumps nevertheless the I-ball solution exists~\cite{Takeda:2014qma}.
Since the Starobinsky model corresponds to the E-model with $\alpha=1$, I-ball solution exist as we have proven. 
In fact, the ref.\cite{Takeda:2014qma} showed the I-balls are formed in the Minkowski spacetime by the numerical simulation. 
In the expanding universe, however, the instability damps so quickly that the fluctuation can not reach the non-linear regime. 
Thus, we must examine whether instability sufficiently grows against the cosmic expansion by performing the linear instability analysis.

To see the growth of the instability, we divide the inflaton field $\phi$ into the background $\phi_0(t)$ and the fluctuation $\delta\phi(x,t)$. Then, the equation of motion for $\phi$
\begin{align}
\ddot{\phi}+3H\dot\phi-\frac1{a^2}\Delta\phi+V'(\phi)=0,
\label{eq:eom}
\end{align}
is decomposed into that for $\phi_0(t)$ and $\delta\phi(x,t)$;
\begin{align}
&\ddot{\phi_0}+3H\dot\phi_0+V'(\phi_0)=0,\\
&\delta\ddot{\phi_k}+3H\delta\dot\phi_k+\left[\frac{k^2}{a^2}+V''(\phi_0)\right]\delta\phi_k=0,\label{eq:fli}
\end{align}
where $a$ is a scale factor and $\delta\phi_k$ are the Fourier modes of $\delta\phi(x)$.
To discuss the evolution of the fluctuations, we first solve for the background $\phi_0$.
In the small field regime $\Phi_0/\sqrt{\alpha}\Mpl\equiv\epsilon_0\lesssim1$, the oscillation is dominated by the quadratic term and we can find the solution perturbatively as in the previous section, which is given as follows:
\begin{align}\nonumber
\phi_0(t)\simeq&\Phi_0\left[\cos(\tau)+\frac{\sqrt{6}}{12}\epsilon_0(3-\cos(2\tau))+\frac1{18}\epsilon_0^2\cos(3\tau)\right]\\
&+\mathcal{O}(\epsilon_0^3),\end{align}
where $\Phi_0$ is a constant and $\tau$ are given as 
\begin{align}
\tau=\sqrt{1-\frac{2}{3}\epsilon_0^2}mt+\mathcal{O}(\epsilon_0^3),
\end{align}
here we temporarily ignore the cosmic expansion.
Plugging this into Eq.~(\ref{eq:fli}), we obtain the following frequency for the fluctuation $\delta\phi_k$:
\begin{align}\nonumber
\omega_k^2\simeq &\frac{k^2}{a^2}+m^2\left[1-\frac13\epsilon_0^2-\sqrt{6}\epsilon_0\cos(\tau)+\frac53\epsilon_0^2\cos(2\tau)\right]\\
&+\mathcal{O}(\epsilon_0^3).
\end{align}
We can see that the resonance is triggered by the linear combination of the last two terms up to the second-order perturbation. 
The leading source of the parametric resonance is the lowest harmonic term $\sim\epsilon_0\cos(\tau)$ which leads to the following Mathieu equation:
\begin{align}
&\delta\phi_k''+[A_k+2q\cos(2T)]\delta\phi_k,\\
&A_k=\frac{k^2}{a^2m^2}\left(4+\frac83\epsilon_0^2\right)+4+\frac43\epsilon_0^2,~q=2\sqrt6\epsilon_0,
\end{align}
where $(')$ represents the derivative with respect to $T\equiv\tau/2$. 
When the dynamics is described in the small field regime, $q\lesssim1$ and the instability is induced by the narrow resonance~\cite{Kofman:1997yn}. 
In this case, the second instability band $-q^2/2\lesssim A_k-4\lesssim q^2/2$ is responsible for the resonance and the modes satisfying
\begin{align}
\frac{k}{m}\lesssim a\frac23\sqrt{6}\epsilon_0
\end{align}
can grow. 

However, in the presence of the cosmic expansion, the amplitude of the background oscillation $\Phi_0$ damps and the instability band gets narrower by factor $a^{-1/2}$. The Floquet index $\mu=q/2$, which represents the growth rate of the instability, also diminishes in time. Since the I-ball formation requires $ t_{\rm form}\sim100/m$, this narrow resonance can not grow the instability sufficiently against the cosmic expansion.

Thus, in the aim of the I-ball formation, broad resonance with $q\gtrsim1$ must take place. Since this requires $\epsilon\gtrsim1$, the maximum value of the $\epsilon$, that is, $\Phi_{\rm end}/\sqrt{\alpha}M_{\rm pl}$ should be larger than unity. According to Eq.~(\ref{pe}), it gives a upper bound on $\alpha$ such that
\begin{align}
\alpha\lesssim 0.84.
\end{align}
This result reproduces the fact the I-balls are not formed in the Starobinsky model $\alpha=1$. 
We note that this is only a necessary condition for the broad resonance and does not guarantee the I-ball formation.

To find the threshold value of $\alpha$ for the I-ball formation precisely, we have numerically solved Eq.~(\ref{eq:fli}) and followed the growth of the instability for various $\alpha$. 
As a result, we find the fluctuations can sufficiently grows and reach the non-linear regime for $\alpha\lesssim 10^{-3}$.
Conversely, the instability is damped for $\alpha\gtrsim 10^{-3}$ due to the cosmic expansion.
We present the snapshots of the evolution of the fluctuations for the choice $\alpha=8\times10^{-4}$.
\begin{figure}[t]
\centering
  \includegraphics[width=1\linewidth]{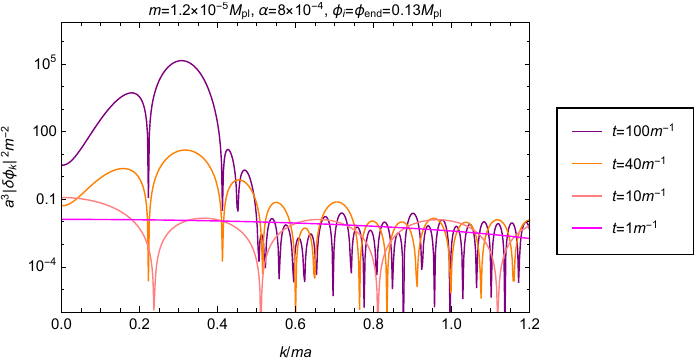}
 \caption{Instability bands for $\alpha=8\times10^{-4},~\phi_i=\phi_{\rm end}=0.13M_{\rm pl}$, which are evaluated at various times. }
\label{fig:eins}
\end{figure}
In the calculation, we set the initial conditions as
\begin{align}
\phi_i&=\phi_{\rm end}(\simeq0.13M_{\rm pl}),\\
\delta\phi_i&=\mathcal{O}(10^{-5})\phi_{\rm end}.
\label{eq:init}
\end{align}

The maximal $\alpha$ for the I-ball formation in the E-models is rather larger than that is predicted in the case of T-models, which is about $10^{-4}$~\cite{Lozanov:2016hid,Kim:2017duj}. 
This difference may originate from the cubic term which induces the lowest harmonic oscillator $\sim\cos(t)$ and is absent in the T-models due to their $Z_2$-symmetric potential.
Furthermore, since the amplitude of the oscillation of the mass is large due to the exponential potential for $\phi<0$, the Floquet index may take a larger value than in the case of T-models.

\section{Lattice simulation of I-balls formation}
\label{sec:nu}
In order to follow the dynamics in non-linear regime, we also performed the lattice simulation of the formation of I-balls using the modified version of CLUSTEREASY~\cite{Felder:2000hq}, which is the parallel computing version of LATTICEEASY, which in turn is a C++ program designed for simulating scalar field evolution in an expanding universe. We integrate the equation of motion Eq.~(\ref{eq:eom}) using the leapfrog method of second order, and approximate the spatial derivatives through the Central-Difference formulas of second order.

The initial value of the scale factor $a$ is normalized as unity, and the Hubble parameter is defined as
\begin{align}
H=\sqrt{\frac{\langle\rho\rangle}{3M_{\rm pl}^2}},
\end{align}
where $\langle~\rangle$ is the average over the lattice and $\rho$ is the energy density, which is given by
\begin{align}
\rho=\frac12\dot{\phi}^2+\frac1{a^2}(\nabla\phi)^2+V(\phi).
\end{align}
In Table~\ref{tab:par}, we present the lattice settings for the simulations.
\begin{table}[t]
\centering
\large
\begin{tabular}{c|ccc}\hline
&{\normalsize $N_{\rm grid}$}&{\normalsize $Lm$}&{\normalsize $m\Delta t$}\\ \hline\hline
{\normalsize1D}&{\normalsize $1024$}&{\normalsize50}&{\normalsize0.04}\\
{\normalsize2D}&{\normalsize$256^2$}&{\normalsize50}&{\normalsize0.1}\\
{\normalsize3D}&{\normalsize$128^3$}&{\normalsize50}&{\normalsize0.2}\\ \hline
  \end{tabular}
 \caption{Lattice settings for simulations.}
\label{tab:par}
\end{table}
The parameters $N_{\rm grid},L$ and $\Delta t$ are the number of grids, box size, and time step, respectively. All the quantities in the program are given as functions of rescaled variables $\phi/\phi_{\rm end},~V/(m\phi_{\rm end})^{2},mt,$ and $mx$, where $x,~t$ are spacetime coordinates. 

\begin{figure}[t]
\centering
  \includegraphics[width=0.85\linewidth]{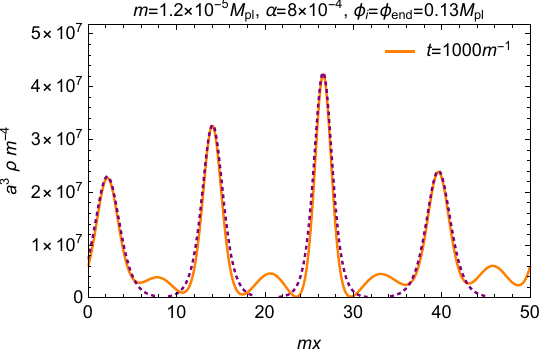}
 \caption{An example of 1D lattice simulations with $\alpha=8\times10^{-4},~\phi_i=\phi_{\rm end}\simeq0.13M_{\rm pl}$, where we set $N_{\rm grid}=1024,~Lm=50$. We plot the comoving energy density at $t=1000m^{-1}$, and fit the peaks to the analytic profiles obtained in Sec.~\ref{sec:bp}, which agree quite well.}
\label{fig:1d}
\end{figure}

In Fig.~\ref{fig:1d}, we present an example of 1D simulations, where we plot energy density $\rho$ after the formation. 

\begin{figure}[t]
\centering
  \includegraphics[width=0.85\linewidth]{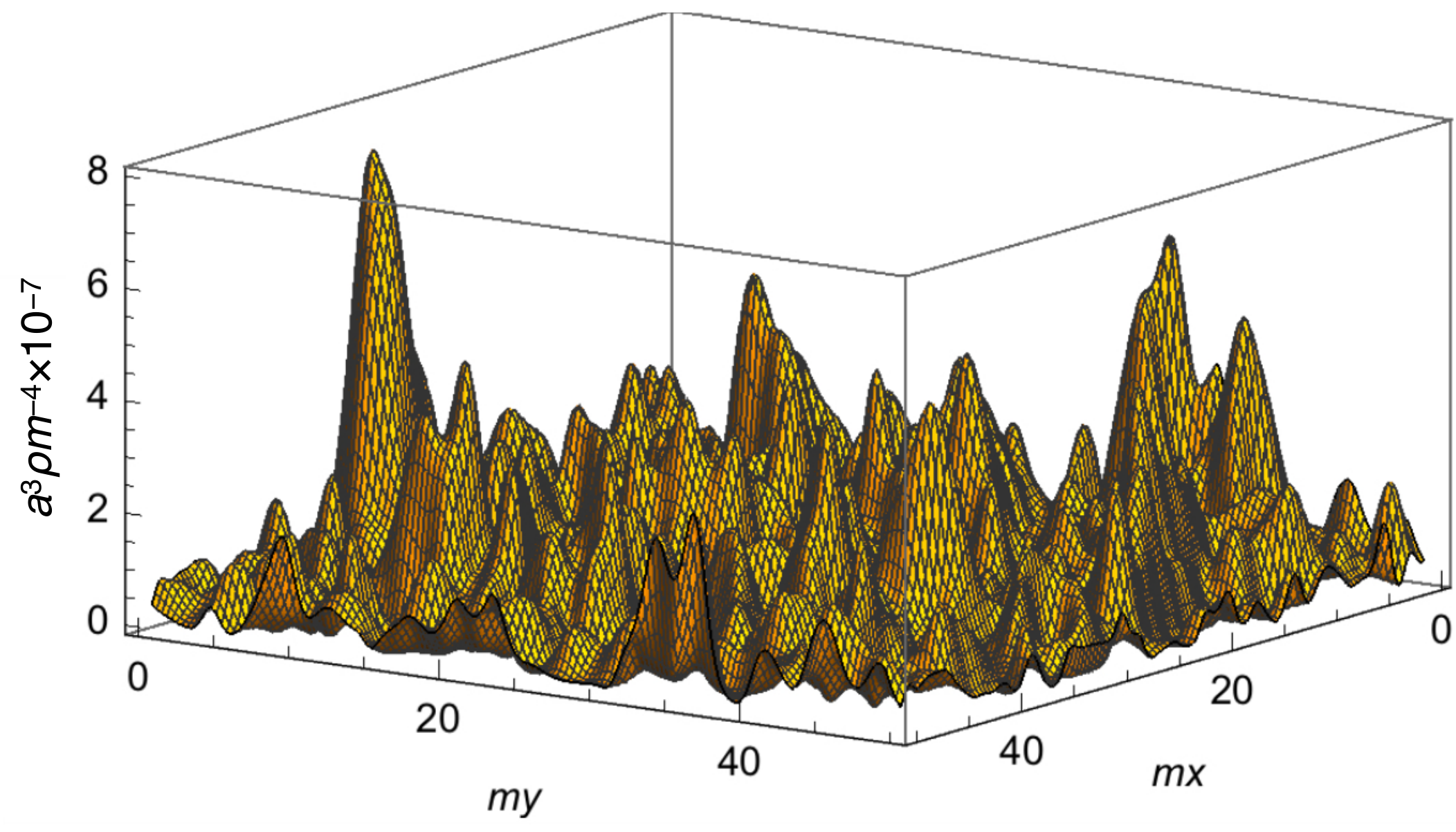}
 \caption{An example of 2D lattice simulation with $\alpha=8\times10^{-4}$ and $\phi_i=\phi_{\rm end}=0.13M_{\rm pl}$, where we set $N_{\rm grid}=256^2,~Lm=50$. We plot the comoving energy density normalized by $10^7m^4$ at $t=2\times10^4m^{-1}$. }
\label{fig:2d}
\end{figure}

\begin{figure}[!h]
\centering
  \includegraphics[width=0.8\linewidth]{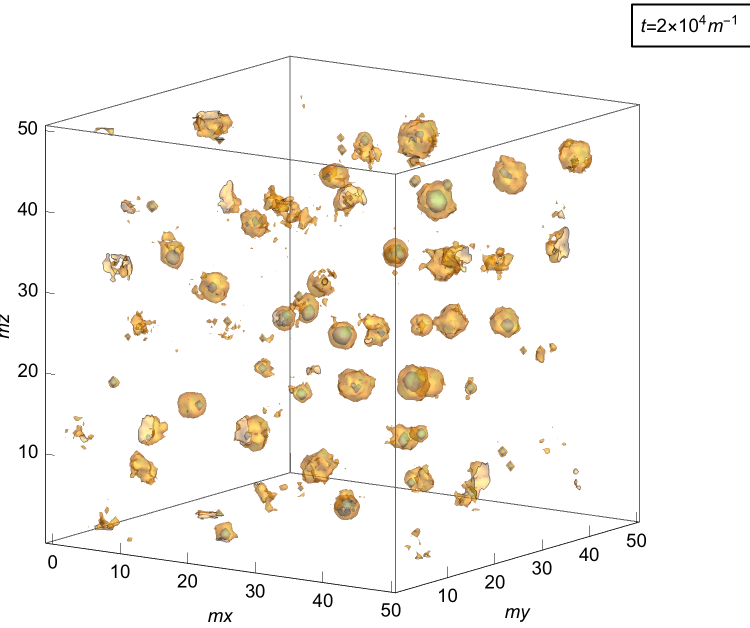}
 \caption{An example of 3D lattice simulation with $\alpha=8\times10^{-4}$ and $\phi_i=\phi_{\rm end}=0.13M_{\rm pl}$, where we set $N_{\rm grid}=128^3,~Lm=50$. We plot the iso-surfaces of the comoving energy density at $a^3\rho=5.9\times10^{7}m^4$ and $a^3\rho=1.8\times10^{8}m^4$ , at the time $t=2\times10^4m^{-1}$.}
\label{fig:3d}
\end{figure}

We found that the formation time is $t_{\rm form}\sim100m^{-1}$. 
We overlapped the energy density of analytic profile Eq.~(\ref{eq:ana2}) with the numerical one in Fig.~\ref{fig:1d}. 
Here we have read the value of $\Phi(0)$, the only one parameter which determines the form of the analytic profile, using $\rho_{\rm peak}=V(\Phi(0))$ and matched the location of the peak.
We can see that both configurations agree well.
In Figures~\ref{fig:2d} and \ref{fig:3d}, we illustrate the examples of 2D and 3D simulations, respectively, where we also plot the energy density after the formation.
In Fig.~\ref{fig:3d}, we plotted iso-surfaces of the energy density at $a^3\rho=5.9\times10^{7}m^4$ and $a^3\rho=1.8\times10^{8}m^4$, where one can clearly see that the localized ball-like objects are actually formed, which we identify as I-balls.   
\section{Cosmological implications of the I-ball formation}
\label{sec:cosmo}
We have confirmed that coherent oscillation of the inflaton fragments into the localized objects called I-balls for small $\alpha$, and estimated the threshold value for the formation as $\alpha_{\rm th}^{\rm E} \sim10^{-3}$.
The fragmentation of the inflaton can have an influence on the cosmology after the inflation because the reheating process is significantly altered. 
In the usual case without inflaton fragmentation, the energy of the inflaton is converted to the light particles mainly by the perturbative decay. 
However, once the inflaton fragments to the I-balls, the universe is dominated by the I-balls and must be reheated by the decay of them. 
Therefore, the time of reheating completion may be changed, unless the perturbative inflaton decay completes before the fragmentation. 
Such situations are studied in the case of a Q-ball, which is a complex scalar configuration that minimizes the energy under the conservation of global $U(1)$ charge~\cite{Enqvist:2002rj,Enqvist:2002si}. Since the center of the Q-ball has a large VEV and matter fields obtain a larger mass, the reheating occurs through the surface effect. 

The I-ball, however, is a real scalar configuration and oscillates at the center as well, hence the reheating occurs over the volume. While there are some studies in particular situations, the general decaying properties of I-balls are actually not known well. There can be self-decay through the self-interactions $\phi^n$~\cite{Hertzberg:2010yz}, and also the decay into other light particles if the proper couplings exist\footnote{The couplings to the other particles do not spoil the I-ball formation as long as the decay is later than the formation time. However, the violent non-perturbative decay into the other particles such as preheating, the inflaton could radiate away within a few oscillations, which is much earlier than the time scale of the I-balls formation.}. 
In particular, it was reported that the decay into the scalar particles is exponentially enhanced due to the Bose stimulation leading to the immediate decay of I-balls after their formation~\cite{Kawasaki:2013awa}. These processes must occur before the BBN.

We note that such a change in the reheating scenario also affects the predictions on the inflationary observables. As we can see in Eqs.~(\ref{amp}),~(\ref{ns}) and (\ref{r}), the prediction of the primordial perturbation depends on $N_*$, which is the $e$-foldings number at CMB pivot scale. Although it is roughly assumed to take a value of $50\sim60$, the actual value of $N_*$ is completely determined by specifying the reheating temperature due to the following relation:
\begin{align}
N_*=61.4+\frac{1}{2}\ln\left(\frac{V_*}{M_{\rm pl}^4}\right)-\frac{1}{3}\ln\left(\frac{\rho_{e}}{M_{\rm pl}^4}\right)+\frac{1}{3}\ln\left(\frac{T_R}{M_{\rm pl}}\right)
\end{align}
Thus, the I-ball formation alters the predictions on the inflationary observables through the reheating temperature. 
For instance, if the decay rate of the I-ball is smaller than the perturbative one, lower reheating temperature realizes, which makes $N_*$ smaller. 
This implies that $N_*$ gets the $\alpha$ dependence due to the I-balls formation for $\alpha<10^{-3}$, which is negligible in absence of I-ball formation. 
On the contrary, if the decay is sufficiently fast, the reheating temperature becomes higher, leading to the larger $N_*$. While large $N_*$ may cause the overproduction of gravitinos in the thermal bath, it is favored by latest Planck observations.

It is worth to comment on the difference between E-models and T-models. 
As we mentioned, the predictions of the two models degenerate for $\alpha\ll N_*$ and the same reheating temperatures. 
However, the conditions for I-ball formation are different: $\alpha^{\rm T}_{\rm th}\sim10^{-4},\alpha^{\rm E}_{\rm th}\sim10^{-3}$. 
This means that there are possibilities that this difference resolves the degeneracy of the inflationary observables through $N_*$. 
The same discussion applies for UV-motivated inflation models which predict tiny tensor mode and degenerate with the $\alpha$-attractors for $\alpha\ll1$, e.g, polyinstanton inflation~($\alpha\sim10^{-3}$)~\cite{Cicoli:2011ct}, K\"{a}hler moduli inflation~($\alpha\sim10^{-8}$)~\cite{Conlon:2005jm}.

\section{Conclusions and discussion}
\label{sec:conc}
Recently proposed $\alpha$-attractors are categorized into two subclasses: T-models and E-models, which are favored by the observations, due to the flatness of the potentials. 
It is known that such a flatness of the potential makes the inflaton fragment into the quasi-stable objects called I-balls.  
In this paper, we investigated the possibility of the formation of I-balls in E-models. 
For small $\alpha$, $\phi$ feels the flatness for longer time, hence the negative pressure may induce the instability and I-balls are formed.
By using the linear instability analysis, and also performing the lattice simulation, we actually confirmed that the I-balls are formed for $\alpha\ltsim\alpha^{\rm E}_{\rm th}\sim10^{-3}$.
The maximal $\alpha$ for the I-balls formation is rather larger than in the case of T-models, which is about $\alpha^{\rm T}_{\rm th}\sim10^{-4}$~\cite{Lozanov:2016hid,Kim:2017duj}. 
This difference would be due to the cubic term in the E-models, which is absent in the T-models. 
The formation of the I-balls after inflation can have an influence on the reheating properties. Since the reheating temperature is altered from the perturbative one, the cosmological scenarios such as baryogenesis, gravitino production are affected. Also the predictions on the inflationary variables $n_s$ and $r$ can be changed, since the CMB pivot scale $N_*$ is related to the reheating temperature.
\vspace{0.2cm}

\noindent{\bf Note added}

While we were finalizing this work, the paper~\cite{Lozanov:2017hjm} by Lozanov, et al. appeared, which discusses the fragmentation of the inflaton field in the cosmological $\alpha$-attractors. Although they focus on the impacts of the short-lived objects which appear in the case of $n>1$, they calculate the instability bands for E-models with $n=1$. Our results in Sec.~\ref{sec:dr} are consistent with their analysis. Moreover, we show the numerical simulations of the I-ball formation and point out the difference in the critical value of $\alpha$ for I-ball formation  between E-models and T-models, which may solve the degeneracy of the inflationary observables predicted by each model.
\section*{Acknowledgments}
The authors would like to thank Masahiro Kawasaki and Naoyuki Takeda for helpful comments. F. H. is supported by JSPS Research Fellowship for Young Scientists Grant Number 17J07391.

\bibliographystyle{apsrev4-1}
\bibliography{IE}

\begin{thebibliography}{39}%
\makeatletter
\providecommand \@ifxundefined [1]{%
 \@ifx{#1\undefined}
}%
\providecommand \@ifnum [1]{%
 \ifnum #1\expandafter \@firstoftwo
 \else \expandafter \@secondoftwo
 \fi
}%
\providecommand \@ifx [1]{%
 \ifx #1\expandafter \@firstoftwo
 \else \expandafter \@secondoftwo
 \fi
}%
\providecommand \natexlab [1]{#1}%
\providecommand \enquote  [1]{``#1''}%
\providecommand \bibnamefont  [1]{#1}%
\providecommand \bibfnamefont [1]{#1}%
\providecommand \citenamefont [1]{#1}%
\providecommand \href@noop [0]{\@secondoftwo}%
\providecommand \href [0]{\begingroup \@sanitize@url \@href}%
\providecommand \@href[1]{\@@startlink{#1}\@@href}%
\providecommand \@@href[1]{\endgroup#1\@@endlink}%
\providecommand \@sanitize@url [0]{\catcode `\\12\catcode `\$12\catcode
  `\&12\catcode `\#12\catcode `\^12\catcode `\_12\catcode `\%12\relax}%
\providecommand \@@startlink[1]{}%
\providecommand \@@endlink[0]{}%
\providecommand \url  [0]{\begingroup\@sanitize@url \@url }%
\providecommand \@url [1]{\endgroup\@href {#1}{\urlprefix }}%
\providecommand \urlprefix  [0]{URL }%
\providecommand \Eprint [0]{\href }%
\providecommand \doibase [0]{http://dx.doi.org/}%
\providecommand \selectlanguage [0]{\@gobble}%
\providecommand \bibinfo  [0]{\@secondoftwo}%
\providecommand \bibfield  [0]{\@secondoftwo}%
\providecommand \translation [1]{[#1]}%
\providecommand \BibitemOpen [0]{}%
\providecommand \bibitemStop [0]{}%
\providecommand \bibitemNoStop [0]{.\EOS\space}%
\providecommand \EOS [0]{\spacefactor3000\relax}%
\providecommand \BibitemShut  [1]{\csname bibitem#1\endcsname}%
\let\auto@bib@innerbib\@empty
\bibitem [{\citenamefont {Starobinsky}(1980)}]{Starobinsky:1980te}%
  \BibitemOpen
  \bibfield  {author} {\bibinfo {author} {\bibfnamefont {A.~A.}\ \bibnamefont
  {Starobinsky}},\ }\href {\doibase 10.1016/0370-2693(80)90670-X} {\bibfield
  {journal} {\bibinfo  {journal} {Phys. Lett.}\ }\textbf {\bibinfo {volume}
  {91B}},\ \bibinfo {pages} {99} (\bibinfo {year} {1980})}\BibitemShut
  {NoStop}%
\bibitem [{\citenamefont {Guth}(1981)}]{Guth:1980zm}%
  \BibitemOpen
  \bibfield  {author} {\bibinfo {author} {\bibfnamefont {A.~H.}\ \bibnamefont
  {Guth}},\ }\href {\doibase 10.1103/PhysRevD.23.347} {\bibfield  {journal}
  {\bibinfo  {journal} {Phys. Rev.}\ }\textbf {\bibinfo {volume} {D23}},\
  \bibinfo {pages} {347} (\bibinfo {year} {1981})}\BibitemShut {NoStop}%
\bibitem [{\citenamefont {Linde}(1982)}]{Linde:1981mu}%
  \BibitemOpen
  \bibfield  {author} {\bibinfo {author} {\bibfnamefont {A.~D.}\ \bibnamefont
  {Linde}},\ }\href {\doibase 10.1016/0370-2693(82)91219-9} {\bibfield
  {journal} {\bibinfo  {journal} {Phys. Lett.}\ }\textbf {\bibinfo {volume}
  {108B}},\ \bibinfo {pages} {389} (\bibinfo {year} {1982})}\BibitemShut
  {NoStop}%
\bibitem [{\citenamefont {Linde}(1983)}]{Linde:1983gd}%
  \BibitemOpen
  \bibfield  {author} {\bibinfo {author} {\bibfnamefont {A.~D.}\ \bibnamefont
  {Linde}},\ }\href {\doibase 10.1016/0370-2693(83)90837-7} {\bibfield
  {journal} {\bibinfo  {journal} {Phys. Lett.}\ }\textbf {\bibinfo {volume}
  {129B}},\ \bibinfo {pages} {177} (\bibinfo {year} {1983})}\BibitemShut
  {NoStop}%
\bibitem [{\citenamefont {Ade}\ \emph {et~al.}(2016)\citenamefont {Ade} \emph
  {et~al.}}]{Ade:2015lrj}%
  \BibitemOpen
  \bibfield  {author} {\bibinfo {author} {\bibfnamefont {P.~A.~R.}\
  \bibnamefont {Ade}} \emph {et~al.} (\bibinfo {collaboration} {Planck}),\
  }\href {\doibase 10.1051/0004-6361/201525898} {\bibfield  {journal} {\bibinfo
   {journal} {Astron. Astrophys.}\ }\textbf {\bibinfo {volume} {594}},\
  \bibinfo {pages} {A20} (\bibinfo {year} {2016})},\ \Eprint
  {http://arxiv.org/abs/1502.02114} {arXiv:1502.02114 [astro-ph.CO]}
  \BibitemShut {NoStop}%
\bibitem [{\citenamefont {Kallosh}\ and\ \citenamefont
  {Linde}(2013)}]{Kallosh:2013hoa}%
  \BibitemOpen
  \bibfield  {author} {\bibinfo {author} {\bibfnamefont {R.}~\bibnamefont
  {Kallosh}}\ and\ \bibinfo {author} {\bibfnamefont {A.}~\bibnamefont
  {Linde}},\ }\href {\doibase 10.1088/1475-7516/2013/07/002} {\bibfield
  {journal} {\bibinfo  {journal} {JCAP}\ }\textbf {\bibinfo {volume} {1307}},\
  \bibinfo {pages} {002} (\bibinfo {year} {2013})},\ \Eprint
  {http://arxiv.org/abs/1306.5220} {arXiv:1306.5220 [hep-th]} \BibitemShut
  {NoStop}%
\bibitem [{\citenamefont {Galante}\ \emph {et~al.}(2015)\citenamefont
  {Galante}, \citenamefont {Kallosh}, \citenamefont {Linde},\ and\
  \citenamefont {Roest}}]{Galante:2014ifa}%
  \BibitemOpen
  \bibfield  {author} {\bibinfo {author} {\bibfnamefont {M.}~\bibnamefont
  {Galante}}, \bibinfo {author} {\bibfnamefont {R.}~\bibnamefont {Kallosh}},
  \bibinfo {author} {\bibfnamefont {A.}~\bibnamefont {Linde}}, \ and\ \bibinfo
  {author} {\bibfnamefont {D.}~\bibnamefont {Roest}},\ }\href {\doibase
  10.1103/PhysRevLett.114.141302} {\bibfield  {journal} {\bibinfo  {journal}
  {Phys. Rev. Lett.}\ }\textbf {\bibinfo {volume} {114}},\ \bibinfo {pages}
  {141302} (\bibinfo {year} {2015})},\ \Eprint {http://arxiv.org/abs/1412.3797}
  {arXiv:1412.3797 [hep-th]} \BibitemShut {NoStop}%
\bibitem [{\citenamefont {Kallosh}\ and\ \citenamefont
  {Linde}(2015)}]{Kallosh:2015lwa}%
  \BibitemOpen
  \bibfield  {author} {\bibinfo {author} {\bibfnamefont {R.}~\bibnamefont
  {Kallosh}}\ and\ \bibinfo {author} {\bibfnamefont {A.}~\bibnamefont
  {Linde}},\ }\href {\doibase 10.1103/PhysRevD.91.083528} {\bibfield  {journal}
  {\bibinfo  {journal} {Phys. Rev.}\ }\textbf {\bibinfo {volume} {D91}},\
  \bibinfo {pages} {083528} (\bibinfo {year} {2015})},\ \Eprint
  {http://arxiv.org/abs/1502.07733} {arXiv:1502.07733 [astro-ph.CO]}
  \BibitemShut {NoStop}%
\bibitem [{\citenamefont {Mukhanov}\ and\ \citenamefont
  {Chibisov}(1981)}]{Mukhanov:1981xt}%
  \BibitemOpen
  \bibfield  {author} {\bibinfo {author} {\bibfnamefont {V.~F.}\ \bibnamefont
  {Mukhanov}}\ and\ \bibinfo {author} {\bibfnamefont {G.~V.}\ \bibnamefont
  {Chibisov}},\ }\href@noop {} {\bibfield  {journal} {\bibinfo  {journal} {JETP
  Lett.}\ }\textbf {\bibinfo {volume} {33}},\ \bibinfo {pages} {532} (\bibinfo
  {year} {1981})}\BibitemShut {NoStop}%
\bibitem [{\citenamefont {Copeland}\ \emph {et~al.}(1995)\citenamefont
  {Copeland}, \citenamefont {Gleiser},\ and\ \citenamefont
  {Muller}}]{Copeland:1995fq}%
  \BibitemOpen
  \bibfield  {author} {\bibinfo {author} {\bibfnamefont {E.~J.}\ \bibnamefont
  {Copeland}}, \bibinfo {author} {\bibfnamefont {M.}~\bibnamefont {Gleiser}}, \
  and\ \bibinfo {author} {\bibfnamefont {H.~R.}\ \bibnamefont {Muller}},\
  }\href {\doibase 10.1103/PhysRevD.52.1920} {\bibfield  {journal} {\bibinfo
  {journal} {Phys. Rev.}\ }\textbf {\bibinfo {volume} {D52}},\ \bibinfo {pages}
  {1920} (\bibinfo {year} {1995})},\ \Eprint
  {http://arxiv.org/abs/hep-ph/9503217} {arXiv:hep-ph/9503217 [hep-ph]}
  \BibitemShut {NoStop}%
\bibitem [{\citenamefont {Kasuya}\ \emph {et~al.}(2003)\citenamefont {Kasuya},
  \citenamefont {Kawasaki},\ and\ \citenamefont {Takahashi}}]{Kasuya:2002zs}%
  \BibitemOpen
  \bibfield  {author} {\bibinfo {author} {\bibfnamefont {S.}~\bibnamefont
  {Kasuya}}, \bibinfo {author} {\bibfnamefont {M.}~\bibnamefont {Kawasaki}}, \
  and\ \bibinfo {author} {\bibfnamefont {F.}~\bibnamefont {Takahashi}},\ }\href
  {\doibase 10.1016/S0370-2693(03)00344-7} {\bibfield  {journal} {\bibinfo
  {journal} {Phys. Lett.}\ }\textbf {\bibinfo {volume} {B559}},\ \bibinfo
  {pages} {99} (\bibinfo {year} {2003})},\ \Eprint
  {http://arxiv.org/abs/hep-ph/0209358} {arXiv:hep-ph/0209358 [hep-ph]}
  \BibitemShut {NoStop}%
\bibitem [{\citenamefont {Amin}\ \emph {et~al.}(2010)\citenamefont {Amin},
  \citenamefont {Easther},\ and\ \citenamefont {Finkel}}]{Amin:2010dc}%
  \BibitemOpen
  \bibfield  {author} {\bibinfo {author} {\bibfnamefont {M.~A.}\ \bibnamefont
  {Amin}}, \bibinfo {author} {\bibfnamefont {R.}~\bibnamefont {Easther}}, \
  and\ \bibinfo {author} {\bibfnamefont {H.}~\bibnamefont {Finkel}},\ }\href
  {\doibase 10.1088/1475-7516/2010/12/001} {\bibfield  {journal} {\bibinfo
  {journal} {JCAP}\ }\textbf {\bibinfo {volume} {1012}},\ \bibinfo {pages}
  {001} (\bibinfo {year} {2010})},\ \Eprint {http://arxiv.org/abs/1009.2505}
  {arXiv:1009.2505 [astro-ph.CO]} \BibitemShut {NoStop}%
\bibitem [{\citenamefont {Gleiser}\ \emph {et~al.}(2011)\citenamefont
  {Gleiser}, \citenamefont {Graham},\ and\ \citenamefont
  {Stamatopoulos}}]{Gleiser:2011xj}%
  \BibitemOpen
  \bibfield  {author} {\bibinfo {author} {\bibfnamefont {M.}~\bibnamefont
  {Gleiser}}, \bibinfo {author} {\bibfnamefont {N.}~\bibnamefont {Graham}}, \
  and\ \bibinfo {author} {\bibfnamefont {N.}~\bibnamefont {Stamatopoulos}},\
  }\href {\doibase 10.1103/PhysRevD.83.096010} {\bibfield  {journal} {\bibinfo
  {journal} {Phys. Rev.}\ }\textbf {\bibinfo {volume} {D83}},\ \bibinfo {pages}
  {096010} (\bibinfo {year} {2011})},\ \Eprint {http://arxiv.org/abs/1103.1911}
  {arXiv:1103.1911 [hep-th]} \BibitemShut {NoStop}%
\bibitem [{\citenamefont {Gleiser}(1994)}]{Gleiser:1993pt}%
  \BibitemOpen
  \bibfield  {author} {\bibinfo {author} {\bibfnamefont {M.}~\bibnamefont
  {Gleiser}},\ }\href {\doibase 10.1103/PhysRevD.49.2978} {\bibfield  {journal}
  {\bibinfo  {journal} {Phys. Rev.}\ }\textbf {\bibinfo {volume} {D49}},\
  \bibinfo {pages} {2978} (\bibinfo {year} {1994})},\ \Eprint
  {http://arxiv.org/abs/hep-ph/9308279} {arXiv:hep-ph/9308279 [hep-ph]}
  \BibitemShut {NoStop}%
\bibitem [{\citenamefont {Fodor}\ \emph {et~al.}(2006)\citenamefont {Fodor},
  \citenamefont {Forgacs}, \citenamefont {Grandclement},\ and\ \citenamefont
  {Racz}}]{Fodor:2006zs}%
  \BibitemOpen
  \bibfield  {author} {\bibinfo {author} {\bibfnamefont {G.}~\bibnamefont
  {Fodor}}, \bibinfo {author} {\bibfnamefont {P.}~\bibnamefont {Forgacs}},
  \bibinfo {author} {\bibfnamefont {P.}~\bibnamefont {Grandclement}}, \ and\
  \bibinfo {author} {\bibfnamefont {I.}~\bibnamefont {Racz}},\ }\href {\doibase
  10.1103/PhysRevD.74.124003} {\bibfield  {journal} {\bibinfo  {journal} {Phys.
  Rev.}\ }\textbf {\bibinfo {volume} {D74}},\ \bibinfo {pages} {124003}
  (\bibinfo {year} {2006})},\ \Eprint {http://arxiv.org/abs/hep-th/0609023}
  {arXiv:hep-th/0609023 [hep-th]} \BibitemShut {NoStop}%
\bibitem [{\citenamefont {Kolb}\ and\ \citenamefont
  {Tkachev}(1994)}]{Kolb:1993hw}%
  \BibitemOpen
  \bibfield  {author} {\bibinfo {author} {\bibfnamefont {E.~W.}\ \bibnamefont
  {Kolb}}\ and\ \bibinfo {author} {\bibfnamefont {I.~I.}\ \bibnamefont
  {Tkachev}},\ }\href {\doibase 10.1103/PhysRevD.49.5040} {\bibfield  {journal}
  {\bibinfo  {journal} {Phys. Rev.}\ }\textbf {\bibinfo {volume} {D49}},\
  \bibinfo {pages} {5040} (\bibinfo {year} {1994})},\ \Eprint
  {http://arxiv.org/abs/astro-ph/9311037} {arXiv:astro-ph/9311037 [astro-ph]}
  \BibitemShut {NoStop}%
\bibitem [{\citenamefont {Antusch}\ \emph {et~al.}(2018)\citenamefont
  {Antusch}, \citenamefont {Cefala}, \citenamefont {Krippendorf}, \citenamefont
  {Muia}, \citenamefont {Orani},\ and\ \citenamefont
  {Quevedo}}]{Antusch:2017flz}%
  \BibitemOpen
  \bibfield  {author} {\bibinfo {author} {\bibfnamefont {S.}~\bibnamefont
  {Antusch}}, \bibinfo {author} {\bibfnamefont {F.}~\bibnamefont {Cefala}},
  \bibinfo {author} {\bibfnamefont {S.}~\bibnamefont {Krippendorf}}, \bibinfo
  {author} {\bibfnamefont {F.}~\bibnamefont {Muia}}, \bibinfo {author}
  {\bibfnamefont {S.}~\bibnamefont {Orani}}, \ and\ \bibinfo {author}
  {\bibfnamefont {F.}~\bibnamefont {Quevedo}},\ }\href {\doibase
  10.1007/JHEP01(2018)083} {\bibfield  {journal} {\bibinfo  {journal} {JHEP}\
  }\textbf {\bibinfo {volume} {01}},\ \bibinfo {pages} {083} (\bibinfo {year}
  {2018})},\ \Eprint {http://arxiv.org/abs/1708.08922} {arXiv:1708.08922
  [hep-th]} \BibitemShut {NoStop}%
\bibitem [{\citenamefont {Mukaida}\ \emph {et~al.}(2017)\citenamefont
  {Mukaida}, \citenamefont {Takimoto},\ and\ \citenamefont
  {Yamada}}]{Mukaida:2016hwd}%
  \BibitemOpen
  \bibfield  {author} {\bibinfo {author} {\bibfnamefont {K.}~\bibnamefont
  {Mukaida}}, \bibinfo {author} {\bibfnamefont {M.}~\bibnamefont {Takimoto}}, \
  and\ \bibinfo {author} {\bibfnamefont {M.}~\bibnamefont {Yamada}},\ }\href
  {\doibase 10.1007/JHEP03(2017)122} {\bibfield  {journal} {\bibinfo  {journal}
  {JHEP}\ }\textbf {\bibinfo {volume} {03}},\ \bibinfo {pages} {122} (\bibinfo
  {year} {2017})},\ \Eprint {http://arxiv.org/abs/1612.07750} {arXiv:1612.07750
  [hep-ph]} \BibitemShut {NoStop}%
\bibitem [{\citenamefont {McDonald}(2002)}]{McDonald:2001iv}%
  \BibitemOpen
  \bibfield  {author} {\bibinfo {author} {\bibfnamefont {J.}~\bibnamefont
  {McDonald}},\ }\href {\doibase 10.1103/PhysRevD.66.043525} {\bibfield
  {journal} {\bibinfo  {journal} {Phys. Rev.}\ }\textbf {\bibinfo {volume}
  {D66}},\ \bibinfo {pages} {043525} (\bibinfo {year} {2002})},\ \Eprint
  {http://arxiv.org/abs/hep-ph/0105235} {arXiv:hep-ph/0105235 [hep-ph]}
  \BibitemShut {NoStop}%
\bibitem [{\citenamefont {Antusch}\ \emph {et~al.}(2017)\citenamefont
  {Antusch}, \citenamefont {Cefala},\ and\ \citenamefont
  {Orani}}]{Antusch:2016con}%
  \BibitemOpen
  \bibfield  {author} {\bibinfo {author} {\bibfnamefont {S.}~\bibnamefont
  {Antusch}}, \bibinfo {author} {\bibfnamefont {F.}~\bibnamefont {Cefala}}, \
  and\ \bibinfo {author} {\bibfnamefont {S.}~\bibnamefont {Orani}},\ }\href
  {\doibase 10.1103/PhysRevLett.118.011303} {\bibfield  {journal} {\bibinfo
  {journal} {Phys. Rev. Lett.}\ }\textbf {\bibinfo {volume} {118}},\ \bibinfo
  {pages} {011303} (\bibinfo {year} {2017})},\ \Eprint
  {http://arxiv.org/abs/1607.01314} {arXiv:1607.01314 [astro-ph.CO]}
  \BibitemShut {NoStop}%
\bibitem [{\citenamefont {Liu}\ \emph {et~al.}(2018)\citenamefont {Liu},
  \citenamefont {Guo}, \citenamefont {Cai},\ and\ \citenamefont
  {Shiu}}]{Liu:2017hua}%
  \BibitemOpen
  \bibfield  {author} {\bibinfo {author} {\bibfnamefont {J.}~\bibnamefont
  {Liu}}, \bibinfo {author} {\bibfnamefont {Z.-K.}\ \bibnamefont {Guo}},
  \bibinfo {author} {\bibfnamefont {R.-G.}\ \bibnamefont {Cai}}, \ and\
  \bibinfo {author} {\bibfnamefont {G.}~\bibnamefont {Shiu}},\ }\href {\doibase
  10.1103/PhysRevLett.120.031301} {\bibfield  {journal} {\bibinfo  {journal}
  {Phys. Rev. Lett.}\ }\textbf {\bibinfo {volume} {120}},\ \bibinfo {pages}
  {031301} (\bibinfo {year} {2018})},\ \Eprint
  {http://arxiv.org/abs/1707.09841} {arXiv:1707.09841 [astro-ph.CO]}
  \BibitemShut {NoStop}%
\bibitem [{\citenamefont {Lozanov}\ and\ \citenamefont
  {Amin}(2017{\natexlab{a}})}]{Lozanov:2016hid}%
  \BibitemOpen
  \bibfield  {author} {\bibinfo {author} {\bibfnamefont {K.~D.}\ \bibnamefont
  {Lozanov}}\ and\ \bibinfo {author} {\bibfnamefont {M.~A.}\ \bibnamefont
  {Amin}},\ }\href {\doibase 10.1103/PhysRevLett.119.061301} {\bibfield
  {journal} {\bibinfo  {journal} {Phys. Rev. Lett.}\ }\textbf {\bibinfo
  {volume} {119}},\ \bibinfo {pages} {061301} (\bibinfo {year}
  {2017}{\natexlab{a}})},\ \Eprint {http://arxiv.org/abs/1608.01213}
  {arXiv:1608.01213 [astro-ph.CO]} \BibitemShut {NoStop}%
\bibitem [{\citenamefont {Kim}\ and\ \citenamefont
  {McDonald}(2017)}]{Kim:2017duj}%
  \BibitemOpen
  \bibfield  {author} {\bibinfo {author} {\bibfnamefont {J.}~\bibnamefont
  {Kim}}\ and\ \bibinfo {author} {\bibfnamefont {J.}~\bibnamefont {McDonald}},\
  }\href {\doibase 10.1103/PhysRevD.95.123537} {\bibfield  {journal} {\bibinfo
  {journal} {Phys. Rev.}\ }\textbf {\bibinfo {volume} {D95}},\ \bibinfo {pages}
  {123537} (\bibinfo {year} {2017})},\ \Eprint
  {http://arxiv.org/abs/1702.08777} {arXiv:1702.08777 [astro-ph.CO]}
  \BibitemShut {NoStop}%
\bibitem [{\citenamefont {Takeda}\ and\ \citenamefont
  {Watanabe}(2014)}]{Takeda:2014qma}%
  \BibitemOpen
  \bibfield  {author} {\bibinfo {author} {\bibfnamefont {N.}~\bibnamefont
  {Takeda}}\ and\ \bibinfo {author} {\bibfnamefont {Y.}~\bibnamefont
  {Watanabe}},\ }\href {\doibase 10.1103/PhysRevD.90.023519} {\bibfield
  {journal} {\bibinfo  {journal} {Phys. Rev.}\ }\textbf {\bibinfo {volume}
  {D90}},\ \bibinfo {pages} {023519} (\bibinfo {year} {2014})},\ \Eprint
  {http://arxiv.org/abs/1405.3830} {arXiv:1405.3830 [astro-ph.CO]} \BibitemShut
  {NoStop}%
\bibitem [{\citenamefont {Carrasco}\ \emph
  {et~al.}(2015{\natexlab{a}})\citenamefont {Carrasco}, \citenamefont
  {Kallosh}, \citenamefont {Linde},\ and\ \citenamefont
  {Roest}}]{Carrasco:2015uma}%
  \BibitemOpen
  \bibfield  {author} {\bibinfo {author} {\bibfnamefont {J.~J.~M.}\
  \bibnamefont {Carrasco}}, \bibinfo {author} {\bibfnamefont {R.}~\bibnamefont
  {Kallosh}}, \bibinfo {author} {\bibfnamefont {A.}~\bibnamefont {Linde}}, \
  and\ \bibinfo {author} {\bibfnamefont {D.}~\bibnamefont {Roest}},\ }\href
  {\doibase 10.1103/PhysRevD.92.041301} {\bibfield  {journal} {\bibinfo
  {journal} {Phys. Rev.}\ }\textbf {\bibinfo {volume} {D92}},\ \bibinfo {pages}
  {041301} (\bibinfo {year} {2015}{\natexlab{a}})},\ \Eprint
  {http://arxiv.org/abs/1504.05557} {arXiv:1504.05557 [hep-th]} \BibitemShut
  {NoStop}%
\bibitem [{\citenamefont {Carrasco}\ \emph
  {et~al.}(2015{\natexlab{b}})\citenamefont {Carrasco}, \citenamefont
  {Kallosh},\ and\ \citenamefont {Linde}}]{Carrasco:2015pla}%
  \BibitemOpen
  \bibfield  {author} {\bibinfo {author} {\bibfnamefont {J.~J.~M.}\
  \bibnamefont {Carrasco}}, \bibinfo {author} {\bibfnamefont {R.}~\bibnamefont
  {Kallosh}}, \ and\ \bibinfo {author} {\bibfnamefont {A.}~\bibnamefont
  {Linde}},\ }\href {\doibase 10.1007/JHEP10(2015)147} {\bibfield  {journal}
  {\bibinfo  {journal} {JHEP}\ }\textbf {\bibinfo {volume} {10}},\ \bibinfo
  {pages} {147} (\bibinfo {year} {2015}{\natexlab{b}})},\ \Eprint
  {http://arxiv.org/abs/1506.01708} {arXiv:1506.01708 [hep-th]} \BibitemShut
  {NoStop}%
\bibitem [{\citenamefont {Carrasco}\ \emph
  {et~al.}(2015{\natexlab{c}})\citenamefont {Carrasco}, \citenamefont
  {Kallosh},\ and\ \citenamefont {Linde}}]{Carrasco:2015rva}%
  \BibitemOpen
  \bibfield  {author} {\bibinfo {author} {\bibfnamefont {J.~J.~M.}\
  \bibnamefont {Carrasco}}, \bibinfo {author} {\bibfnamefont {R.}~\bibnamefont
  {Kallosh}}, \ and\ \bibinfo {author} {\bibfnamefont {A.}~\bibnamefont
  {Linde}},\ }\href {\doibase 10.1103/PhysRevD.92.063519} {\bibfield  {journal}
  {\bibinfo  {journal} {Phys. Rev.}\ }\textbf {\bibinfo {volume} {D92}},\
  \bibinfo {pages} {063519} (\bibinfo {year} {2015}{\natexlab{c}})},\ \Eprint
  {http://arxiv.org/abs/1506.00936} {arXiv:1506.00936 [hep-th]} \BibitemShut
  {NoStop}%
\bibitem [{\citenamefont {Farhi}\ \emph {et~al.}(2008)\citenamefont {Farhi},
  \citenamefont {Graham}, \citenamefont {Guth}, \citenamefont {Iqbal},
  \citenamefont {Rosales},\ and\ \citenamefont {Stamatopoulos}}]{Farhi:2007wj}%
  \BibitemOpen
  \bibfield  {author} {\bibinfo {author} {\bibfnamefont {E.}~\bibnamefont
  {Farhi}}, \bibinfo {author} {\bibfnamefont {N.}~\bibnamefont {Graham}},
  \bibinfo {author} {\bibfnamefont {A.~H.}\ \bibnamefont {Guth}}, \bibinfo
  {author} {\bibfnamefont {N.}~\bibnamefont {Iqbal}}, \bibinfo {author}
  {\bibfnamefont {R.~R.}\ \bibnamefont {Rosales}}, \ and\ \bibinfo {author}
  {\bibfnamefont {N.}~\bibnamefont {Stamatopoulos}},\ }\href {\doibase
  10.1103/PhysRevD.77.085019} {\bibfield  {journal} {\bibinfo  {journal} {Phys.
  Rev.}\ }\textbf {\bibinfo {volume} {D77}},\ \bibinfo {pages} {085019}
  (\bibinfo {year} {2008})},\ \Eprint {http://arxiv.org/abs/0712.3034}
  {arXiv:0712.3034 [hep-th]} \BibitemShut {NoStop}%
\bibitem [{\citenamefont {Fodor}\ \emph {et~al.}(2009)\citenamefont {Fodor},
  \citenamefont {Forgacs}, \citenamefont {Horvath},\ and\ \citenamefont
  {Mezei}}]{Fodor:2009kf}%
  \BibitemOpen
  \bibfield  {author} {\bibinfo {author} {\bibfnamefont {G.}~\bibnamefont
  {Fodor}}, \bibinfo {author} {\bibfnamefont {P.}~\bibnamefont {Forgacs}},
  \bibinfo {author} {\bibfnamefont {Z.}~\bibnamefont {Horvath}}, \ and\
  \bibinfo {author} {\bibfnamefont {M.}~\bibnamefont {Mezei}},\ }\href
  {\doibase 10.1016/j.physletb.2009.03.054} {\bibfield  {journal} {\bibinfo
  {journal} {Phys. Lett.}\ }\textbf {\bibinfo {volume} {B674}},\ \bibinfo
  {pages} {319} (\bibinfo {year} {2009})},\ \Eprint
  {http://arxiv.org/abs/0903.0953} {arXiv:0903.0953 [hep-th]} \BibitemShut
  {NoStop}%
\bibitem [{\citenamefont {Dashen}\ \emph {et~al.}(1975)\citenamefont {Dashen},
  \citenamefont {Hasslacher},\ and\ \citenamefont {Neveu}}]{Dashen:1975hd}%
  \BibitemOpen
  \bibfield  {author} {\bibinfo {author} {\bibfnamefont {R.~F.}\ \bibnamefont
  {Dashen}}, \bibinfo {author} {\bibfnamefont {B.}~\bibnamefont {Hasslacher}},
  \ and\ \bibinfo {author} {\bibfnamefont {A.}~\bibnamefont {Neveu}},\ }\href
  {\doibase 10.1103/PhysRevD.11.3424} {\bibfield  {journal} {\bibinfo
  {journal} {Phys. Rev.}\ }\textbf {\bibinfo {volume} {D11}},\ \bibinfo {pages}
  {3424} (\bibinfo {year} {1975})}\BibitemShut {NoStop}%
\bibitem [{\citenamefont {Kofman}\ \emph {et~al.}(1997)\citenamefont {Kofman},
  \citenamefont {Linde},\ and\ \citenamefont {Starobinsky}}]{Kofman:1997yn}%
  \BibitemOpen
  \bibfield  {author} {\bibinfo {author} {\bibfnamefont {L.}~\bibnamefont
  {Kofman}}, \bibinfo {author} {\bibfnamefont {A.~D.}\ \bibnamefont {Linde}}, \
  and\ \bibinfo {author} {\bibfnamefont {A.~A.}\ \bibnamefont {Starobinsky}},\
  }\href {\doibase 10.1103/PhysRevD.56.3258} {\bibfield  {journal} {\bibinfo
  {journal} {Phys. Rev.}\ }\textbf {\bibinfo {volume} {D56}},\ \bibinfo {pages}
  {3258} (\bibinfo {year} {1997})},\ \Eprint
  {http://arxiv.org/abs/hep-ph/9704452} {arXiv:hep-ph/9704452 [hep-ph]}
  \BibitemShut {NoStop}%
\bibitem [{\citenamefont {Felder}\ and\ \citenamefont
  {Tkachev}(2008)}]{Felder:2000hq}%
  \BibitemOpen
  \bibfield  {author} {\bibinfo {author} {\bibfnamefont {G.~N.}\ \bibnamefont
  {Felder}}\ and\ \bibinfo {author} {\bibfnamefont {I.}~\bibnamefont
  {Tkachev}},\ }\href {\doibase 10.1016/j.cpc.2008.02.009} {\bibfield
  {journal} {\bibinfo  {journal} {Comput. Phys. Commun.}\ }\textbf {\bibinfo
  {volume} {178}},\ \bibinfo {pages} {929} (\bibinfo {year} {2008})},\ \Eprint
  {http://arxiv.org/abs/hep-ph/0011159} {arXiv:hep-ph/0011159 [hep-ph]}
  \BibitemShut {NoStop}%
\bibitem [{\citenamefont {Enqvist}\ \emph
  {et~al.}(2002{\natexlab{a}})\citenamefont {Enqvist}, \citenamefont {Kasuya},\
  and\ \citenamefont {Mazumdar}}]{Enqvist:2002rj}%
  \BibitemOpen
  \bibfield  {author} {\bibinfo {author} {\bibfnamefont {K.}~\bibnamefont
  {Enqvist}}, \bibinfo {author} {\bibfnamefont {S.}~\bibnamefont {Kasuya}}, \
  and\ \bibinfo {author} {\bibfnamefont {A.}~\bibnamefont {Mazumdar}},\ }\href
  {\doibase 10.1103/PhysRevLett.89.091301} {\bibfield  {journal} {\bibinfo
  {journal} {Phys. Rev. Lett.}\ }\textbf {\bibinfo {volume} {89}},\ \bibinfo
  {pages} {091301} (\bibinfo {year} {2002}{\natexlab{a}})},\ \Eprint
  {http://arxiv.org/abs/hep-ph/0204270} {arXiv:hep-ph/0204270 [hep-ph]}
  \BibitemShut {NoStop}%
\bibitem [{\citenamefont {Enqvist}\ \emph
  {et~al.}(2002{\natexlab{b}})\citenamefont {Enqvist}, \citenamefont {Kasuya},\
  and\ \citenamefont {Mazumdar}}]{Enqvist:2002si}%
  \BibitemOpen
  \bibfield  {author} {\bibinfo {author} {\bibfnamefont {K.}~\bibnamefont
  {Enqvist}}, \bibinfo {author} {\bibfnamefont {S.}~\bibnamefont {Kasuya}}, \
  and\ \bibinfo {author} {\bibfnamefont {A.}~\bibnamefont {Mazumdar}},\ }\href
  {\doibase 10.1103/PhysRevD.66.043505} {\bibfield  {journal} {\bibinfo
  {journal} {Phys. Rev.}\ }\textbf {\bibinfo {volume} {D66}},\ \bibinfo {pages}
  {043505} (\bibinfo {year} {2002}{\natexlab{b}})},\ \Eprint
  {http://arxiv.org/abs/hep-ph/0206272} {arXiv:hep-ph/0206272 [hep-ph]}
  \BibitemShut {NoStop}%
\bibitem [{\citenamefont {Hertzberg}(2010)}]{Hertzberg:2010yz}%
  \BibitemOpen
  \bibfield  {author} {\bibinfo {author} {\bibfnamefont {M.~P.}\ \bibnamefont
  {Hertzberg}},\ }\href {\doibase 10.1103/PhysRevD.82.045022} {\bibfield
  {journal} {\bibinfo  {journal} {Phys. Rev.}\ }\textbf {\bibinfo {volume}
  {D82}},\ \bibinfo {pages} {045022} (\bibinfo {year} {2010})},\ \Eprint
  {http://arxiv.org/abs/1003.3459} {arXiv:1003.3459 [hep-th]} \BibitemShut
  {NoStop}%
\bibitem [{\citenamefont {Kawasaki}\ and\ \citenamefont
  {Yamada}(2014)}]{Kawasaki:2013awa}%
  \BibitemOpen
  \bibfield  {author} {\bibinfo {author} {\bibfnamefont {M.}~\bibnamefont
  {Kawasaki}}\ and\ \bibinfo {author} {\bibfnamefont {M.}~\bibnamefont
  {Yamada}},\ }\href {\doibase 10.1088/1475-7516/2014/02/001} {\bibfield
  {journal} {\bibinfo  {journal} {JCAP}\ }\textbf {\bibinfo {volume} {1402}},\
  \bibinfo {pages} {001} (\bibinfo {year} {2014})},\ \Eprint
  {http://arxiv.org/abs/1311.0985} {arXiv:1311.0985 [hep-ph]} \BibitemShut
  {NoStop}%
\bibitem [{\citenamefont {Cicoli}\ \emph {et~al.}(2011)\citenamefont {Cicoli},
  \citenamefont {Pedro},\ and\ \citenamefont {Tasinato}}]{Cicoli:2011ct}%
  \BibitemOpen
  \bibfield  {author} {\bibinfo {author} {\bibfnamefont {M.}~\bibnamefont
  {Cicoli}}, \bibinfo {author} {\bibfnamefont {F.~G.}\ \bibnamefont {Pedro}}, \
  and\ \bibinfo {author} {\bibfnamefont {G.}~\bibnamefont {Tasinato}},\ }\href
  {\doibase 10.1088/1475-7516/2011/12/022} {\bibfield  {journal} {\bibinfo
  {journal} {JCAP}\ }\textbf {\bibinfo {volume} {1112}},\ \bibinfo {pages}
  {022} (\bibinfo {year} {2011})},\ \Eprint {http://arxiv.org/abs/1110.6182}
  {arXiv:1110.6182 [hep-th]} \BibitemShut {NoStop}%
\bibitem [{\citenamefont {Conlon}\ and\ \citenamefont
  {Quevedo}(2006)}]{Conlon:2005jm}%
  \BibitemOpen
  \bibfield  {author} {\bibinfo {author} {\bibfnamefont {J.~P.}\ \bibnamefont
  {Conlon}}\ and\ \bibinfo {author} {\bibfnamefont {F.}~\bibnamefont
  {Quevedo}},\ }\href {\doibase 10.1088/1126-6708/2006/01/146} {\bibfield
  {journal} {\bibinfo  {journal} {JHEP}\ }\textbf {\bibinfo {volume} {01}},\
  \bibinfo {pages} {146} (\bibinfo {year} {2006})},\ \Eprint
  {http://arxiv.org/abs/hep-th/0509012} {arXiv:hep-th/0509012 [hep-th]}
  \BibitemShut {NoStop}%
\bibitem [{\citenamefont {Lozanov}\ and\ \citenamefont
  {Amin}(2017{\natexlab{b}})}]{Lozanov:2017hjm}%
  \BibitemOpen
  \bibfield  {author} {\bibinfo {author} {\bibfnamefont {K.~D.}\ \bibnamefont
  {Lozanov}}\ and\ \bibinfo {author} {\bibfnamefont {M.~A.}\ \bibnamefont
  {Amin}},\ }\href@noop {} {\  (\bibinfo {year} {2017}{\natexlab{b}})},\
  \Eprint {http://arxiv.org/abs/1710.06851} {arXiv:1710.06851 [astro-ph.CO]}
  \BibitemShut {NoStop}%
\end{thebibliography}%

\end{document}